\documentclass[%
preprint,
 amsmath,amssymb,
 aps,
]{revtex4-1}

\usepackage{graphicx}
\usepackage{dcolumn}
\usepackage{bm}
\usepackage{units}
\usepackage[mathlines]{lineno}
\usepackage{subfigure}

\newcommand{\minerva}{\mbox{MINERvA}}

\newcommand{\numu}[0]{\nu_{\mu}}
\newcommand{\nuebar}[0]{\overline{\nu}_{e}}

\newcommand{\nue}[0]{\nu_{e}}

\begin{document}

\preprint{APS/123-QED}

\title{ Evidence for neutral-current diffractive $\pi^{0}$ production from hydrogen in neutrino interactions on hydrocarbon}
\newcommand{\deceased}{Deceased}


\newcommand{\Rutgers}{Rutgers, The State University of New Jersey, Piscataway, New Jersey 08854, USA}
\newcommand{\Hampton}{Hampton University, Dept. of Physics, Hampton, VA 23668, USA}
\newcommand{\Dortmund}{Institute of Physics, Dortmund University, 44221, Germany }
\newcommand{\Otterbein}{Department of Physics, Otterbein University, 1 South Grove Street, Westerville, OH, 43081 USA}
\newcommand{\JMU}{James Madison University, Harrisonburg, Virginia 22807, USA}
\newcommand{\Florida}{University of Florida, Department of Physics, Gainesville, FL 32611}
\newcommand{\UCIrvine}{Department of Physics and Astronomy, University of California, Irvine, Irvine, California 92697-4575, USA}
\newcommand{\CBPF}{Centro Brasileiro de Pesquisas F\'{i}sicas, Rua Dr. Xavier Sigaud 150, Urca, Rio de Janeiro, Rio de Janeiro, 22290-180, Brazil}
\newcommand{\PUCP}{Secci\'{o}n F\'{i}sica, Departamento de Ciencias, Pontificia Universidad Cat\'{o}lica del Per\'{u}, Apartado 1761, Lima, Per\'{u}}
\newcommand{\INRM}{Institute for Nuclear Research of the Russian Academy of Sciences, 117312 Moscow, Russia}
\newcommand{\Jlab}{Jefferson Lab, 12000 Jefferson Avenue, Newport News, VA 23606, USA}
\newcommand{\Pittsburgh}{Department of Physics and Astronomy, University of Pittsburgh, Pittsburgh, Pennsylvania 15260, USA}
\newcommand{\Guanajuato}{Campus Le\'{o}n y Campus Guanajuato, Universidad de Guanajuato, Lascurain de Retana No. 5, Colonia Centro, Guanajuato 36000, Guanajuato M\'{e}xico.}
\newcommand{\Athens}{Department of Physics, University of Athens, GR-15771 Athens, Greece}
\newcommand{\Tufts}{Physics Department, Tufts University, Medford, Massachusetts 02155, USA}
\newcommand{\WM}{Department of Physics, College of William \& Mary, Williamsburg, Virginia 23187, USA}
\newcommand{\FNAL}{Fermi National Accelerator Laboratory, Batavia, Illinois 60510, USA}
\newcommand{\Purdue}{Department of Chemistry and Physics, Purdue University Calumet, Hammond, Indiana 46323, USA}
\newcommand{\MCLA}{Massachusetts College of Liberal Arts, 375 Church Street, North Adams, MA 01247}
\newcommand{\UMD}{Department of Physics, University of Minnesota -- Duluth, Duluth, Minnesota 55812, USA}
\newcommand{\Northwestern}{Northwestern University, Evanston, Illinois 60208}
\newcommand{\UNI}{Universidad Nacional de Ingenier\'{i}a, Apartado 31139, Lima, Per\'{u}}
\newcommand{\Rochester}{University of Rochester, Rochester, New York 14627 USA}
\newcommand{\Austin}{Department of Physics, University of Texas, 1 University Station, Austin, Texas 78712, USA}
\newcommand{\USM}{Departamento de F\'{i}sica, Universidad T\'{e}cnica Federico Santa Mar\'{i}a, Avenida Espa\~{n}a 1680 Casilla 110-V, Valpara\'{i}so, Chile}
\newcommand{\Geneva}{University of Geneva, 1211 Geneva 4, Switzerland}
\newcommand{\Chicago}{Enrico Fermi Institute, University of Chicago, Chicago, IL 60637 USA}
\newcommand{\hired}{}
\newcommand{\OregonState}{Department of Physics, Oregon State University, Corvallis, Oregon 97331, USA}
\newcommand{\oxford}{}
\newcommand{\bmeThanks}{now at SLAC National Accelerator Laboratory, Stanford, CA 94309, USA}
\newcommand{\higueraThanks}{now at University of Houston, Houston, TX 77204, USA}
\newcommand{\damartinezThanks}{now at Illinois Institute of Technology, Chicago, IL 60616, USA}
\newcommand{\mcgivernThanks}{now at Iowa State University, Ames, IA 50011, USA}
\newcommand{\joelmousseauThanks}{now at University of Michigan, Ann Arbor, MI 48109, USA}
\newcommand{\LazaThanks}{also at Department of Physics, University of Antananarivo, Madagascar}
\newcommand{\twaltonThanks}{now at Fermi National Accelerator Laboratory, Batavia, IL 60510, USA}
\newcommand{\jwolcottThanks}{now at Tufts University, Medford, MA 02155, USA}

\author{J.~Wolcott}                       \affiliation{\Rochester} \affiliation{\Tufts}
\author{L.~Aliaga}                        \affiliation{\WM}
\author{O.~Altinok}                       \affiliation{\Tufts}
\author{A.~Bercellie}                     \affiliation{\Rochester}
\author{M.~Betancourt}                    \affiliation{\FNAL}
\author{A.~Bodek}                         \affiliation{\Rochester}
\author{A.~Bravar}                        \affiliation{\Geneva}
\author{H.~Budd}                          \affiliation{\Rochester}
\author{T.~Cai}                           \affiliation{\Rochester}
\author{M.F.~Carneiro}                    \affiliation{\CBPF}
\author{J.~Chvojka}                       \affiliation{\Rochester}
\author{H.~da~Motta}                      \affiliation{\CBPF}
\author{J.~Devan}                         \affiliation{\WM}
\author{S.A.~Dytman}                      \affiliation{\Pittsburgh}
\author{G.A.~D\'{i}az~}                   \affiliation{\Rochester}  \affiliation{\PUCP}
\author{B.~Eberly}\thanks{\bmeThanks}     \affiliation{\Pittsburgh}
\author{E.~Endress}                       \affiliation{\PUCP}
\author{J.~Felix}                         \affiliation{\Guanajuato}
\author{L.~Fields}                        \affiliation{\FNAL}  \affiliation{\Northwestern}
\author{R.~Fine}                          \affiliation{\Rochester}
\author{R.Galindo}                        \affiliation{\USM}
\author{H.~Gallagher}                     \affiliation{\Tufts}
\author{T.~Golan}                         \affiliation{\Rochester}  \affiliation{\FNAL}
\author{R.~Gran}                          \affiliation{\UMD}
\author{D.A.~Harris}                      \affiliation{\FNAL}
\author{A.~Higuera}\thanks{\higueraThanks}  \affiliation{\Rochester}  \affiliation{\Guanajuato}
\author{K.~Hurtado}                       \affiliation{\CBPF}  \affiliation{\UNI}
\author{M.~Kiveni}                        \affiliation{\FNAL}
\author{J.~Kleykamp}                      \affiliation{\Rochester}
\author{M.~Kordosky}                      \affiliation{\WM}
\author{T.~Le}                            \affiliation{\Tufts}  \affiliation{\Rutgers}
\author{E.~Maher}                         \affiliation{\MCLA}
\author{S.~Manly}                         \affiliation{\Rochester}
\author{W.A.~Mann}                        \affiliation{\Tufts}
\author{C.M.~Marshall}                    \affiliation{\Rochester}
\author{D.A.~Martinez~Caicedo}\thanks{\damartinezThanks}  \affiliation{\CBPF}
\author{K.S.~McFarland}                   \affiliation{\Rochester}  \affiliation{\FNAL}
\author{C.L.~McGivern}\thanks{\mcgivernThanks}  \affiliation{\Pittsburgh}
\author{A.M.~McGowan}                     \affiliation{\Rochester}
\author{B.~Messerly}                      \affiliation{\Pittsburgh}
\author{J.~Miller}                        \affiliation{\USM}
\author{A.~Mislivec}                      \affiliation{\Rochester}
\author{J.G.~Morf\'{i}n}                  \affiliation{\FNAL}
\author{J.~Mousseau}\thanks{\joelmousseauThanks}  \affiliation{\Florida}
\author{D.~Naples}                        \affiliation{\Pittsburgh}
\author{J.K.~Nelson}                      \affiliation{\WM}
\author{A.~Norrick}                       \affiliation{\WM}
\author{Nuruzzaman}                       \affiliation{\Rutgers}  \affiliation{\USM}
\author{J.~Osta}                          \affiliation{\FNAL}
\author{V.~Paolone}                       \affiliation{\Pittsburgh}
\author{J.~Park}                          \affiliation{\Rochester}
\author{C.E.~Patrick}                     \affiliation{\Northwestern}
\author{G.N.~Perdue}                      \affiliation{\FNAL}  \affiliation{\Rochester}
\author{L.~Rakotondravohitra}\thanks{\LazaThanks}  \affiliation{\FNAL}
\author{M.A.~Ramirez}                     \affiliation{\Guanajuato}
\author{H.~Ray}                           \affiliation{\Florida}
\author{L.~Ren}                           \affiliation{\Pittsburgh}
\author{D.~Rimal}                         \affiliation{\Florida}
\author{P.A.~Rodrigues}                   \affiliation{\Rochester}
\author{D.~Ruterbories}                   \affiliation{\Rochester}
\author{H.~Schellman}                     \affiliation{\OregonState}  \affiliation{\Northwestern}
\author{D.W.~Schmitz}                     \affiliation{\Chicago}  \affiliation{\FNAL}
\author{C.J.~Solano~Salinas}              \affiliation{\UNI}
\author{S.~S\'{a}nchez~Falero}            \affiliation{\PUCP}
\author{N.~Tagg}                          \affiliation{\Otterbein}
\author{B.G.~Tice}                        \affiliation{\Rutgers}
\author{E.~Valencia}                      \affiliation{\Guanajuato}
\author{T.~Walton}\thanks{\twaltonThanks}  \affiliation{\Hampton}
\author{M.Wospakrik}                      \affiliation{\Florida}
\author{D.~Zhang}                         \affiliation{\WM}
%

\collaboration{\minerva\  Collaboration}\ \noaffiliation

\date{\today}


\begin{abstract}

The \minerva\ experiment observes an excess of  events containing electromagnetic showers relative to the expectation from Monte Carlo simulations in neutral-current neutrino interactions with mean beam energy of \unit[4.5]{GeV} on a hydrocarbon target.  
The excess is characterized and found to be consistent with neutral-current $\pi^{0}$  production with a broad energy distribution peaking at 7~GeV and a total cross section of $0.26 \pm 0.02 (stat) \pm 0.08 (sys)\times 10^{-39}$~cm$^{2}$.  The angular distribution, electromagnetic shower energy, and spatial distribution of the energy depositions of the excess are consistent with expectations from neutrino neutral-current diffractive $\pi^{0}$ production from hydrogen in the hydrocarbon target.  These data comprise the first direct experimental observation and constraint for a reaction that poses an important background process in neutrino oscillation experiments searching for $\numu$ to $\nue$ oscillations.

\begin{description}
\item[PACS numbers]13.15.+g,25.30.Pt
\end{description}
\end{abstract}

\maketitle


\section{\label{Intro}Introduction}

Current and future accelerator-based neutrino-oscillation experiments aim to make high precision measurements of oscillation parameters by examining the $\nue$ and $\nuebar$ content of their beams as a function of neutrino energy in the sub-GeV to few-GeV range\cite{NOvATDR,T2KNIM,dune,SBN,hyperk}.  The  signature of a $\nue$($\nuebar$) charged-current (CC) interaction, the signal in such experiments, is the presence of an e$^{-}$(e$^{+}$) in the final state that originates from the neutrino interaction vertex.  In order to extract the desired parameters, it is necessary to compare the observed signal to a simulation containing all processes that can produce a real single e$^{-}$(e$^{+}$) in the final state as well as processes that can mimic this signature.  Precise estimates of the parameters therefore require accurate and complete models of all potential background processes.  Consequently, it is important to characterize and understand any observations of neutrino-induced events in the sub-GeV to many-GeV range that contain electromagnetic showers. 

In a separate paper, the \minerva\ collaboration reported a measurement of $\nue$ CC quasielastic and quasielastic-like scattering in the NuMI beam at an average neutrino energy of 3.6~GeV \cite{minervaNueCCQE}. During the data analysis leading to those results, an unexpectedly large number of events was observed containing electromagnetic showers likely caused by photon conversions.  In this Letter, this excess of events is measured relative to the expectation based on a sample of simulated data produced using current state-of-the-art models of neutrino production and interactions. These events are seen to exhibit features expected of neutral current (NC) diffractive  $\pi^{0}$ production from hydrogen in the hydrocarbon target. 

These results constitute the first direct experimental observation and characterization of this process.  An analogous process that happens exclusively on nuclei heavier than hydrogen, NC coherent $\pi^{0}$ production, has been observed previously\cite{Faissner1983,Isiksal1984,Bergsma1985,Grabosch1986,Baltay1986,miniboone2008,nomad2009,sciboone2010}; however, the contribution from NC diffractive scattering from hydrogen, when present in the target, has been considered only inclusively with the scattering from the heavier nuclei and not examined separately as is done here.
This measurement offers an experimental constraint on models of NC diffractive $\pi^{0}$ production and the A-dependence of coherent scattering. It is of general interest in neutrino physics and of particular importance for oscillation experiments using detectors containing water or hydrocarbons or any other material containing hydrogen.

\section{\label{ExptData}The \minerva\ experiment}

The \minerva\ experiment studies neutrinos produced in the NuMI
beamline at Fermi National Accelerator Laboratory. This analysis uses data taken between March 2010 and April 2012 with $3.49\times 10^{20}$ protons on target (POT).\cite{NUMIref} During this period, the beam consisted predominantly of $\numu$ with a peak energy of \unit[3.15]{GeV} and a high-energy tail extending up to tens of GeV such that the mean neutrino energy was \unit[4.5]{GeV}.  $\nue$ and $\nuebar$ made up approximately 1.6\% of the total neutrino flux.  The neutrino beam simulation used by \minerva\ is described in Ref. \cite{minervaNueCCQE} and references therein.

The \minerva\ detector \cite{MINERvA_NIM,MINERVA_DAQ} consists of a core of scintillator strips arranged in planes and oriented in three views for three-dimensional tracking.  The triangular strips ($3.4~{\rm cm\ base} \times 1.7~{\rm cm\ height}$) making up the sensitive portion of the detectors are sufficiently fine-grained  
to ensure reliable detection and characterization of electromagnetic showers at energies of above roughly \unit[0.5]{GeV}.  The scintillator core is augmented by electromagnetic and hadronic calorimetry on both the sides and the downstream end of the detector.
The \minerva\ detector's response is simulated by tuned GEANT4-based\cite{Agostinelli,Allison} software.  The energy scale is set by requiring that the photostatistics and reconstructed energy for energy deposited by momentum-analyzed muons traversing the detector agree in data and simulation.  Additional algorithm-specific tuning, including corrections for passive material, is done using the simulation \cite{MINERvA_NIM}.

Simulated neutrino interactions, generated with the GENIE 2.6.2 neutrino event generator \cite{GENIE262}, are used for comparison to the data and efficiency corrections.  Of particular interest in this Letter are processes that contain electromagnetic showers.  The dominant source of electromagnetic showers in these neutrino interactions is neutral pion production, which is modeled in the generator via resonant production from nucleons according to the Rein-Sehgal model; via coherent interactions with nuclei according to the PCAC formalism of Rein and Sehgal \cite{RS}; and via the hadronization model in non-resonant inelastic production.  Further details on other processes simulated by the generator, as well as the external data sets used for tuning the generator, are described briefly in Ref. \cite{minervaNueCCQE} and references therein.

\section{\label{reconanal}Event reconstruction and analysis}

Events of interest were selected as part of the $\nue$ CC quasielastic scattering analysis\cite{minervaNueCCQE}\cite{JWthesis}.  
Candidate events are created from reconstructed tracks originating in the central scintillator region of the \minerva\ detector\cite{MINERvA_NIM}. To remove the overwhelming background from $\nu_{\mu}$ CC events, tracks are not considered if they exit the back of the detector as muons are expected to do.  Candidate electromagnetic showers are identified by examining  energy depositions within a region that consists of the union of two volumes:  a cylinder of  radius \unit[50]{mm} extending from the event vertex along the track direction and a 
$7.5^{\circ}$ cone with an apex at the event vertex (origin of track) and a symmetry axis along the track direction.  
The full region (referred to below as the `shower cone') extends in length through the scintillator and electromagnetic calorimeter portions of \minerva\ until it reaches a gap of approximately three radiation lengths along the cone where no significant energy is deposited.  This shower cone object is evaluated using a multivariate particle identification (PID) algorithm which combines details of the energy deposition pattern both longitudinally (mean $dE/dx$ and the fraction of energy at the downstream end of cone) and transverse to the axis of the cone (mean shower width) using a $k$-nearest-neighbors (kNN) algorithm\cite{knnref}.

\begin{figure}[h]
			\centering
			 \includegraphics[width=0.7\textwidth]{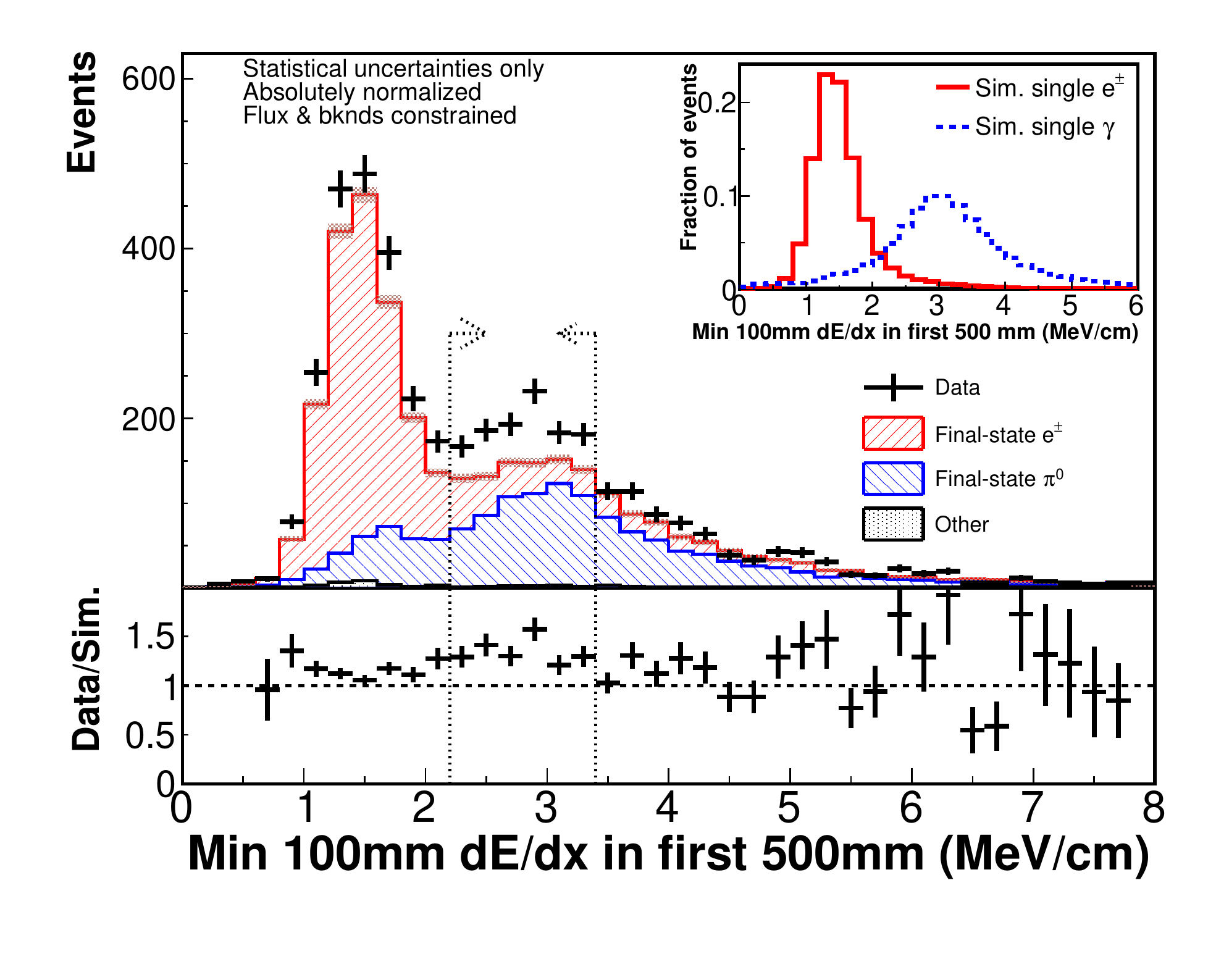}
			\caption{Measure of the minimum $dE/dx$ near the front of candidate electromagnetic showers for data (crosses) and the simulated neutrino event sample (solid).  Simulated events are divided according to the progenitor of the electromagnetic shower.  The dashed lines and arrows delineate the excess region discussed in the text.  Shown at the bottom is the ratio of data to simulation. The inset shows the same distribution for simulated single-particle samples of electrons and photons. }
			\label{fig_dedx}
		\end{figure}

For events deemed by the PID algorithm to be electromagnetic-like, the $dE/dx$ at the front of the shower cone is examined to see if it is more consistent with a single particle, such as that expected from an electron (or positron), or two particles, as would be seen in a photon pair conversion into $e^{+} e^{-}$. Here, the energy in the $dE/dx$ measure is taken to be the minimum energy contained in a \unit[100]{mm} window along the shower, where the downstream end of the window is allowed to slide up to \unit[500]{mm} from the vertex.
This sliding window technique reduces any potential bias induced by nuclear activity near the interaction point\cite{numuPRL}. Figure~\ref{fig_dedx} shows the minimum $dE/dx$ during this process for both the data and simulation.  For comparison,  the inset of Fig.~\ref{fig_dedx} shows the same variable for simulated samples of single photons or electrons, chosen with a flat energy distribution in the range from 0.5 to 5.0 GeV.  Electron showers tend to lie in an interval between 1 and \unit[2]{MeV/cm}, while the photons populate a somewhat wider range peaking at \unit[3]{MeV/cm}.   
The \minerva \ modeling of photons and electrons was validated against the data successfully with samples of separated $\pi^{0}$ conversion photons and Michel electrons\cite{MINERvA_NIM}.  

The electron region of Fig.~\ref{fig_dedx}, peaking at approximately \unit[1.3]{MeV/cm} in both the data and the simulation, is well-modeled; in both shape and magnitude, the data and simulation differ by less than 10\%.  However, the photon peak in the data contains an excess relative to the prediction with $12.5\sigma$ statistical-only significance.  Systematic uncertainties, particularly those associated with the flux model and the estimate of the other processes predicted in that region, reduce the significance to  $3.1\sigma$.  (The overall flux prediction and uncertainties, as well as the normalization of the background processes and corresponding uncertainties for the simulation shown in Fig.~\ref{fig_dedx}, were constrained by \textit{in situ} measurements in dedicated samples.  Both of these and other systematic errors are described in detail in Reference \cite{minervaNueCCQE}.)  

Since distributions made using a sideband sample of $\nue$ events containing Michel electrons agree very well with the simulation, the excess of data events is unlikely to have arisen from the misreconstruction of electrons or errors in the modeling of electromagnetic showers in the simulation.  In addition, Fig.~\ref{fig_dedx} shows that the excess is not compatible with an overall normalization offset of the sample.  The possibility of the excess arising from mismodeled non-shower activity near the event vertex (i.e., nucleons) was examined by injecting extra protons into simulated electron showers in a fashion consistent with the findings in recent \minerva\ muon neutrino scattering results \cite{numuPRL} (uniformly from 0-225 MeV in 25\% of the simulated showers).  These samples did not exhibit an excess in the photon region of the reconstructed $dE/dx$ distribution.   Moreover, as will be shown in the following sections, the excess events in the photon region are qualitatively different than any of the event types predicted by the simulation under the photon peak.  

\section{\label{characteriz}Characterization of the excess}

In order to characterize the excess in Fig.~\ref{fig_dedx}, events exhibiting minimum $dE/dx$ between 2.2 and $\unit[3.4]{MeV/cm}$ were selected in both the data and the simulation.  Kinematic distributions of the candidate EM shower in these events were examined after subtracting the simulation from the data bin-by-bin.  Distributions made in this fashion provide a picture of what is missing in the simulation and thereby characterize the excess.  

\begin{figure*}
	\hspace{-0.015\textwidth}
	\subfigure[]{
		\includegraphics[width=0.48\textwidth]{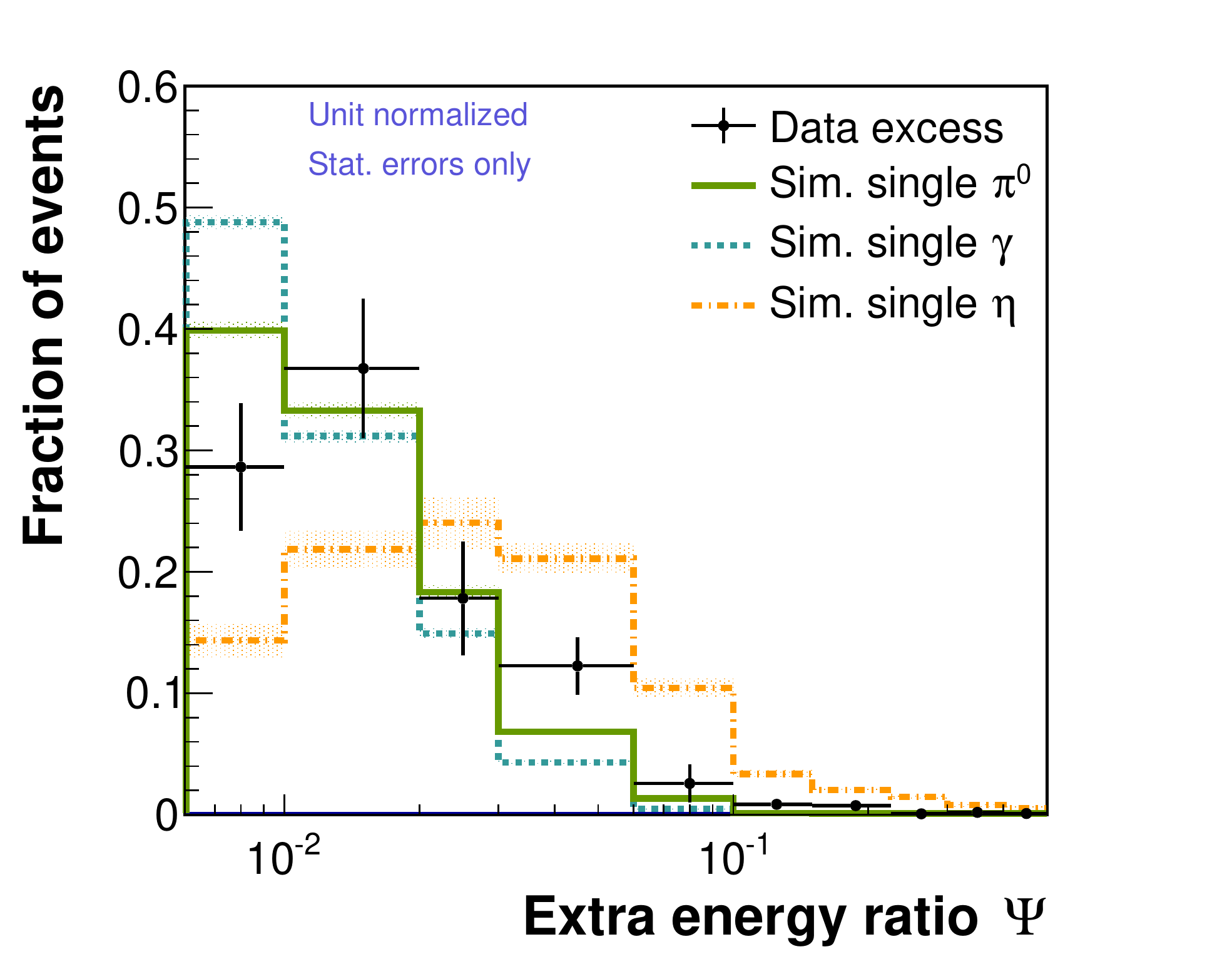}
		\label{fig:psi_PC}
	}
	\subfigure[]{
		\includegraphics[width=0.48\textwidth]{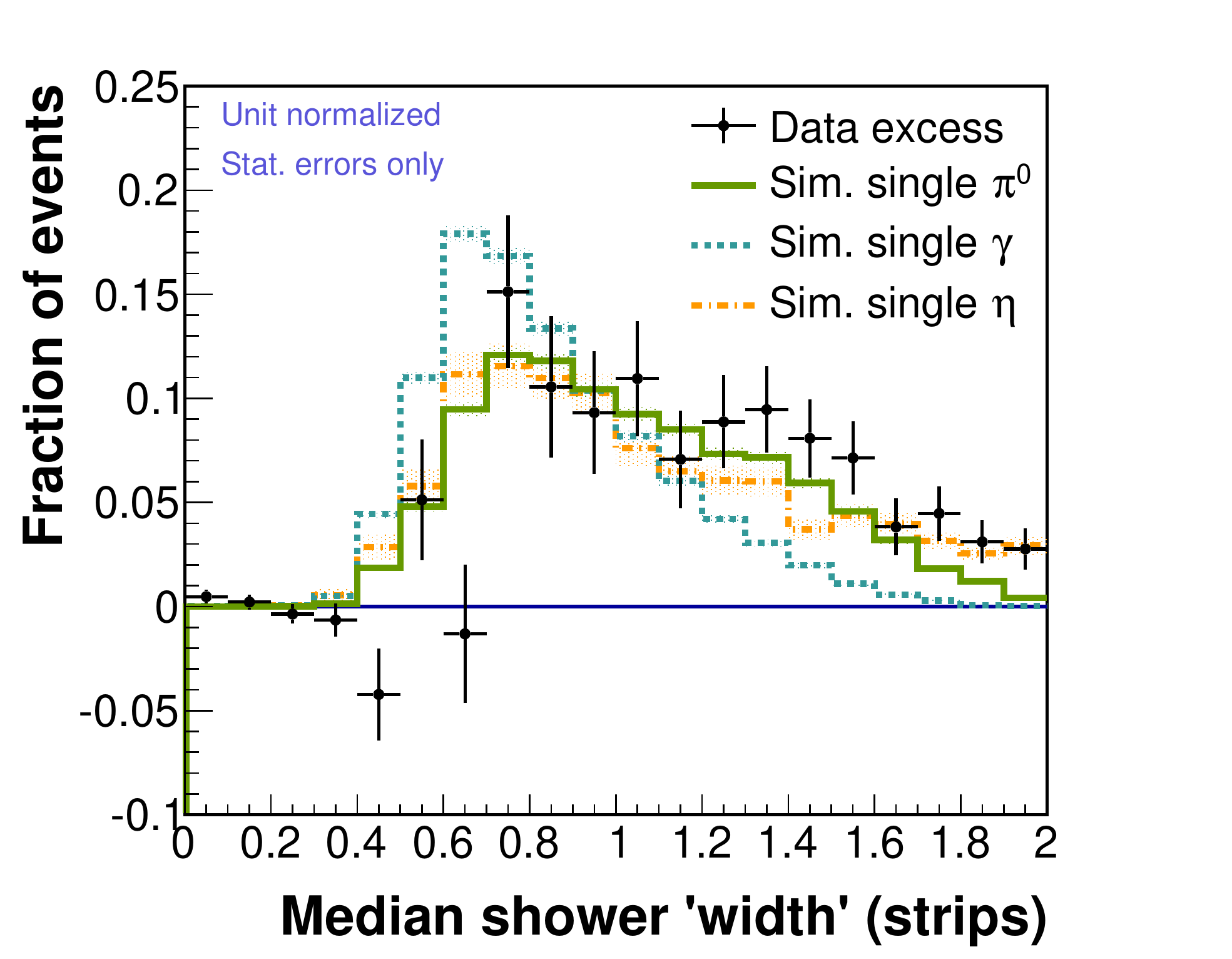}
		\label{fig:MSPW_PC}
	}
	\caption{Left: ratio of energy outside the shower cone to that inside the shower cone for the data excess (points) compared via shape to samples (histograms) created using different single-particle simulations, weighted to have kinematics similar to the excess events. Right: Distributions of median transverse width of the EM shower for the same samples.  Uncertainties are statistical only.}
	\label{EER_parts}
\end{figure*}

The excess was compared to single-particle samples of photons and $\pi^{0}$'s which were simulated with broad distributions in energy (0-20 GeV) and angle with respect to the longitudinal detector axis (0-$\pi$/2) and processed using the \minerva\ reconstruction.  A similar sample of $\eta$'s was also constructed to investigate the possibility of a heavier state decaying into showering particles.  In each of these samples, the events falling into the region of the photon-like excess in $dE/dx$ were generated to have the same two dimensional distribution of energy and angle as in the data excess.
Figure~\ref{fig:psi_PC} shows a shape comparison for the ``extra energy ratio" variable $\Psi$, which represents the relative amount of energy outside the  cone  to that inside the cone, for these single-particle samples compared to the distribution of the excess in the data.   Energy depositions within \unit[30]{cm} of the interaction vertex were ignored when calculating $\Psi$ to reduce the contribution from low-energy nucleons, which may not be simulated correctly \cite{numuPRL}. Here, the data are more consistent with photon or $\pi^{0}$ production than $\eta$ production.  On the other hand, Fig.~\ref{fig:MSPW_PC} shows the median transverse width of the energy depositions in the cone object (``median shower width'') for the single-particle samples and the excess in the data.  In this case, the data are less consistent with the behavior expected from a single particle than  with that from a particle decaying into multiple photons.  These single-particle studies along with the Michel and injected proton studies mentioned above, suggest that the showers in the excess are most likely caused by photons from $\pi^{0}$ production and subsequent decay.    

\begin{figure*}
	\subfigure[]{
		\includegraphics[width=0.48\textwidth]{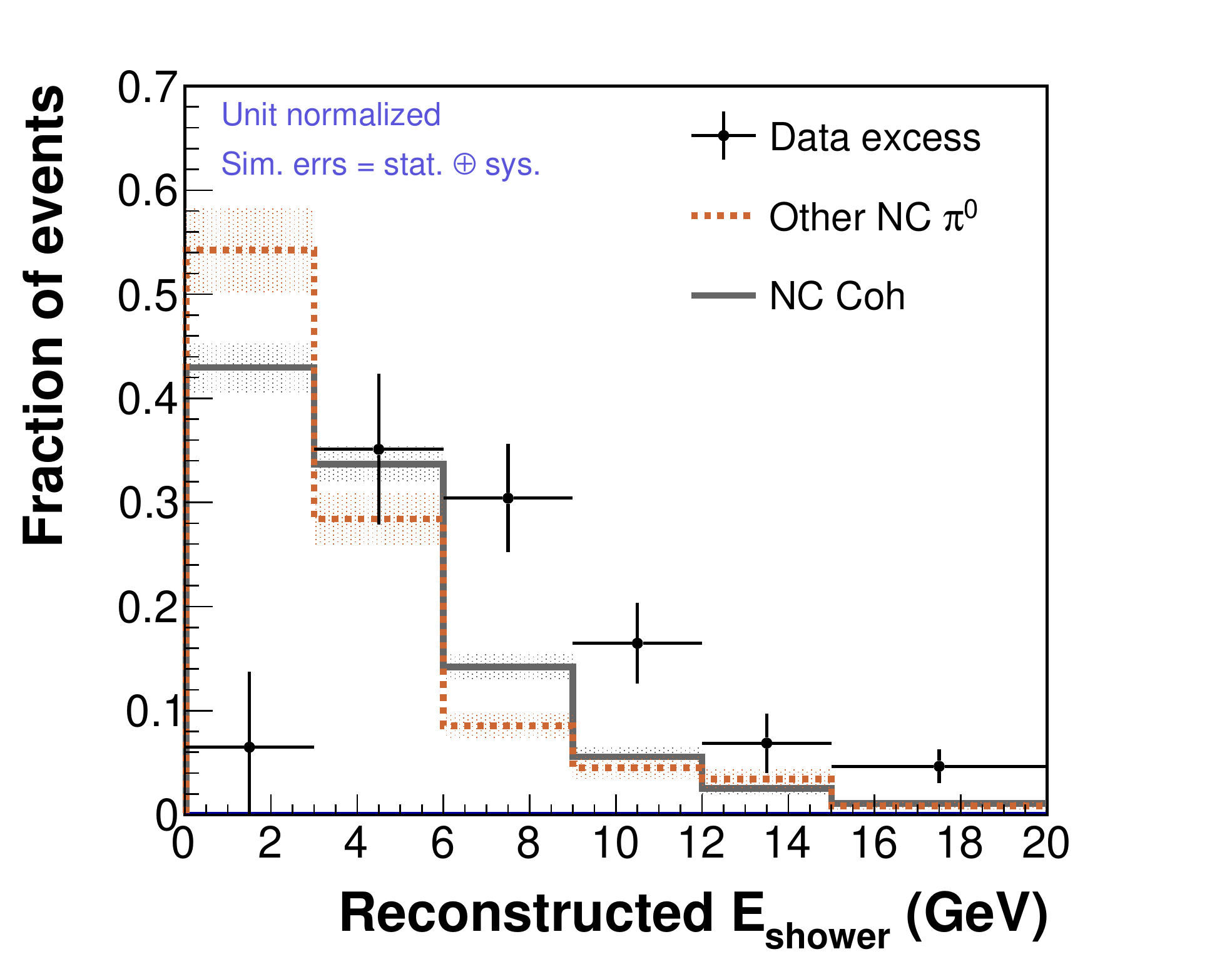}
		\label{fig:Eshower_NC}
	}
	\subfigure[]{
		\includegraphics[width=0.48\textwidth]{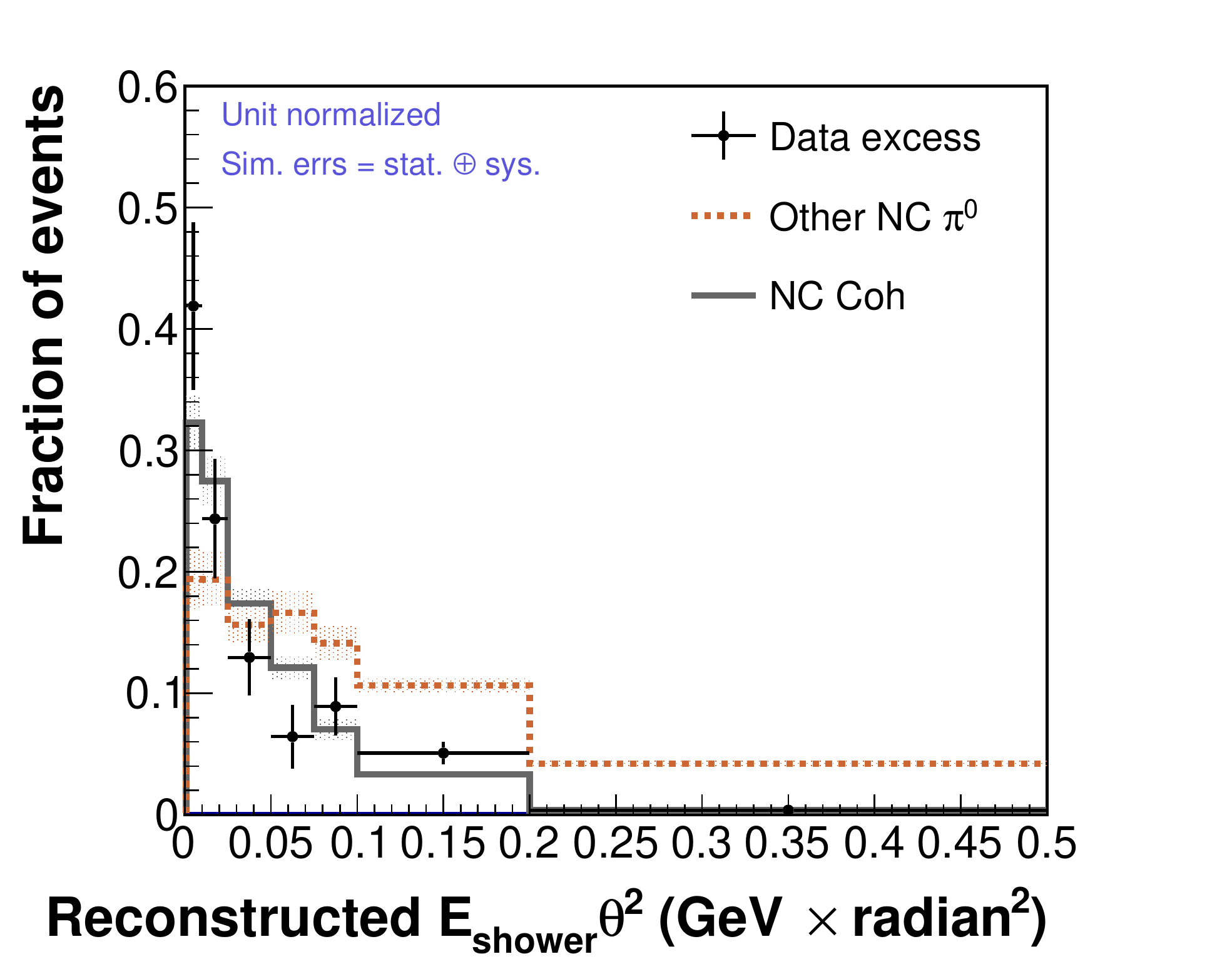}
		\label{fig:ETheta2_NC}
	}
	\subfigure[]{
		\includegraphics[width=0.48\textwidth]{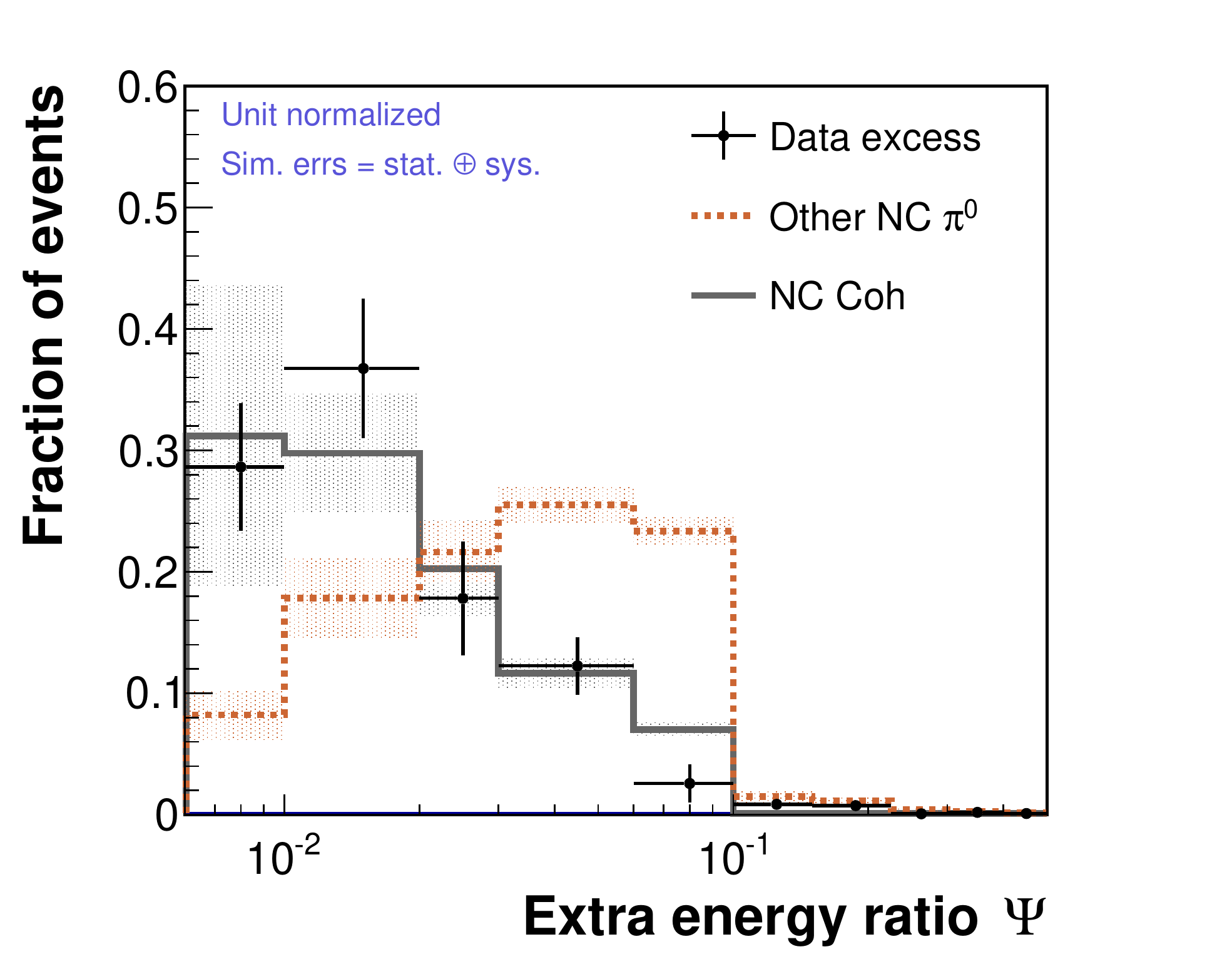}
		\label{fig:Psi_NC}
	}
	\subfigure[]{
		\includegraphics[width=0.48\textwidth]{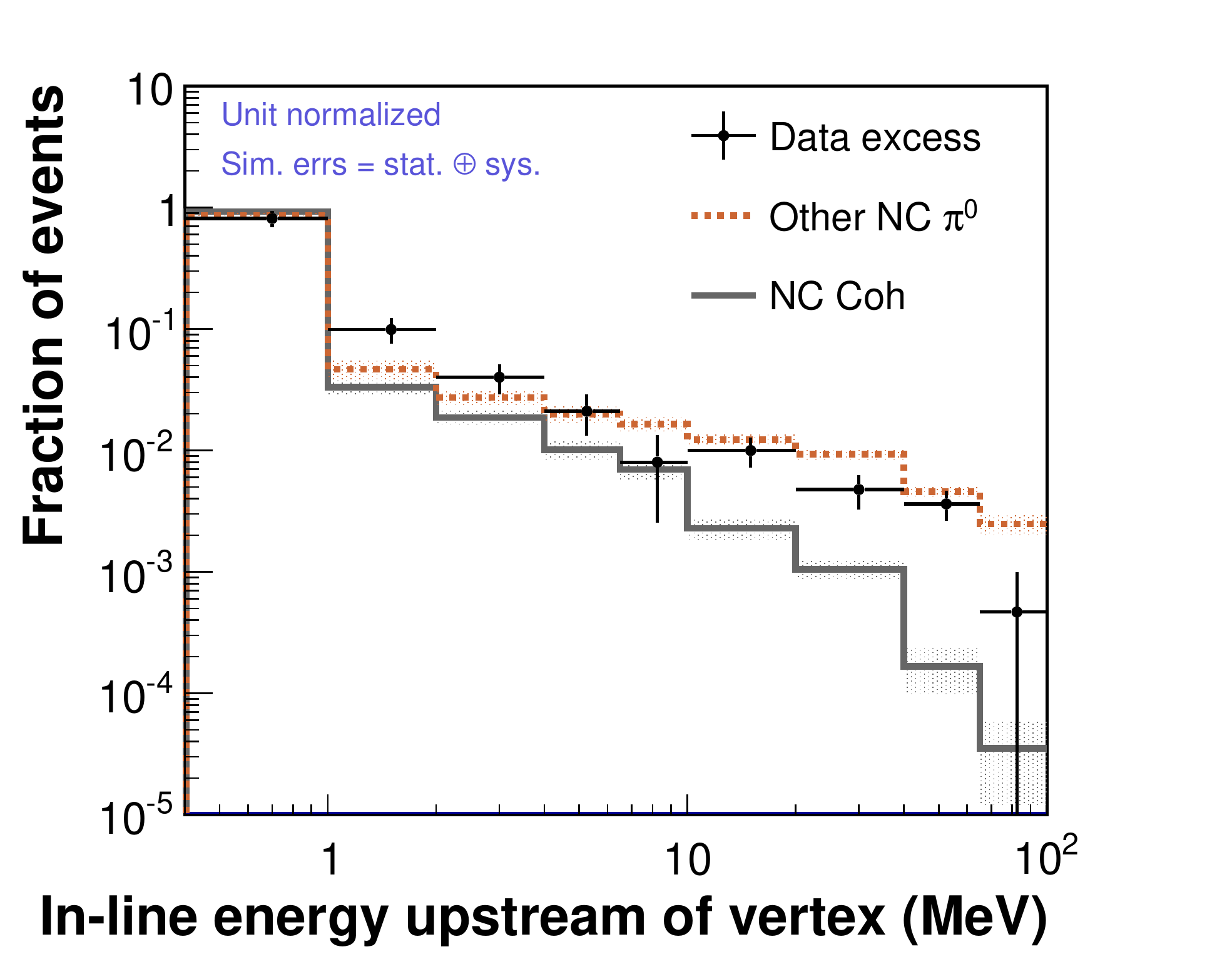}
		\label{fig:UIE_NC}
	}
	\caption{The data excess (points) as compared (via shape) to GENIE samples of NC coherent and incoherent $\pi^{0}$ production. The comparisons are made as a function of $E_{\mathrm{shower}}$ (upper left), $E_{\mathrm{shower}}\theta ^{2}$ (upper right), $\Psi$ (lower left), and in-line upstream energy (lower right).  Data uncertainties are statistcal only; predictions include systematic uncertaintes added in quadrature with statistical.}
	\label{char}
\end{figure*}

The lack of a muon and the fact that the shower exhibits photon-like, rather than electron-like, energy deposition behavior together imply that the process contributing to the excess is a neutral-current (NC) interaction.  Other features of the sample can be examined to provide further insight into the nature of the interaction.  Figure~\ref{char} shows shape comparisons of GENIE NC coherent and incoherent $\pi^{0}$ production with  data distributions from the excess in several variables, where the content of each curve is normalized to unity.  Figure~\ref{fig:Eshower_NC} gives the reconstructed energy the electromagnetic shower, $E_{\mathrm{shower}}$, where it can be seen that the data excess has a harder energy spectrum than the NC processes predicted by the model. However, the angular distribution of the shower in the data agrees very well in shape with the expectation from GENIE for NC coherent $\pi^{0}$ production, as demonstrated in the $E_{\mathrm{shower}}\theta^{2}$ distribution (Fig.~\ref{fig:ETheta2_NC}).  The same is true in $\Psi$, as illustrated in Fig.~\ref{fig:Psi_NC},  as most of the events have relatively little energy outside the cone.  However, the distribution of energy within a cone identical to the one described in Sec. \ref{reconanal}, except oriented backward along the original cone axis, is different. In this case, illustrated in Fig.~\ref{fig:UIE_NC}, the data appear to have more in-line upstream energy than the NC coherent process and are more consistent with the NC incoherent process, suggesting a small nuclear recoil from the neutrino interaction.  Corroborating this hypothesis, the charge-weighted distance from that energy to the shower vertex was examined in the data sample and seen to fit the exponential decay distance expected for a photon conversion after propagating through the detector from the interaction point defined by the upstream activity.

The results described above were supplemented by a visual scan of event displays for data in the excess region and a simulated neutrino event sample, as well as simulated single-particle samples.  The conclusions from the scan were that the data in this region, relative to the simulated sample, contains a higher fraction of events with a $\pi^{0}$ and more events with in-line upstream energy.  

Finally, the difference between the data and the expectation from GENIE between 2.2 and 3.4 MeV/cm in Fig.~\ref{fig_dedx} was used to extract a total cross section for $E_{\mathrm{shower}} > 3~GeV$ integrated over the \minerva\ flux of $0.26 \pm 0.02 (stat) \pm 0.08 (sys) \times 10^{-39} cm^{2}/CH$. The phase space for this measurement was limited in $E_{\mathrm{shower}}$ to avoid model dependence by ensuring the value reported is in a region where \minerva\ has good sensitivity.

\section{\label{source}Diffractive $\pi^{0}$ production}
The most plausible source of the excess seen in the data is diffractive NC $\pi^{0}$ production from hydrogen in the scintillator target of \minerva .  Because little momentum is tranferred to the nucleus, this process is expected to be characterized by coherent-like kinematics; but the comparatively small mass of the hydrogen nucleus would result in the proton sometimes being endowed with sufficient kinetic energy to manifest as in-line upstream energy in this analysis.  In addition, NC diffractive scattering from hydrogen is not included in the GENIE simulation used by \minerva .  

Though neutral-current excitation of a $\Delta^{+}$ from a proton within a nucleus produces the same final state after the decay $\Delta^{+} \rightarrow p + \pi^{0}$, the latter process is characterized by a strong peak around \unit[1.2]{GeV} in the invariant mass spectrum of the events.  The invariant mass distribution for the excess was computed, using the upstream inline energy distribution to form a rough estimate for the proton kinetic energy, and was found to occupy a broad $W$ spectrum peaking at about \unit[3.5]{GeV} with FWHM of about \unit[3]{GeV}.  Thus a deficiency in the resonant production model in GENIE, which simulates this process, is unlikely to be responsible for the excess, and leaves diffractive scattering as the best hypothesis.

To further test the hypothesis that the observed signal arises from diffractive NC $\pi^{0}$ production, comparisons were made to an early implementation in GENIE of a calculation of the diffractive process based on the work of Rein \cite{Rein1986} that is valid for W$>$2.0~GeV.  
This model produces events with a similar cross section to the value observed by \minerva\ for the excess and it contributes events in the region of the excess and very little outside that region.  The model qualitatively agrees with the characteristics of the excess in terms of the shower angle, extra energy ratio and in-line upstream energy (Figs. 3(b), 3(c), and 3(d), respectively), but exhibits a somewhat different shape in terms of the energy spectrum of the produced shower.
Further details of the comparison of the observed excess and the Rein model can be found in Ref. \cite{JWthesis}.

\section{\label{concl}Conclusions}

An excess of events containing electromagnetic showers observed by the \minerva\ experiment appears to originate from the neutral-current production of neutral pions in a process not predicted by the GENIE neutrino interaction simulation program.  Interpretations of the excess as arising from errors in the flux or background predictions, or mismodeling of the electromagnetic shower process, are disfavored based on \textit{in situ} sideband constraints.  The observed process resembles coherent production apart from the existence of a small amount of upstream energy, implying that the events likely arise from diffractive pion production from hydrogen.  The measured cross section  for this process for $E_{\pi}>3~GeV$, assuming the observed shower to come from photon conversions from the $\pi^{0}$, is comparable to that for NC coherent $\pi^{0}$ production from carbon.  These measurements, interpreted as NC diffractive scattering, constitute the first direct experimental observation and characterization of this process.  Neutrino oscillation experiments with hydrogen in their targets must account for NC diffractive scattering in order to correctly model backgrounds to $\nue$ appearance.  The data presented above will play an essential role in constraining models for diffractive production, such as the model in Ref. \cite{Rein1986}.  But because the latter applies only at larger W, this work also highlights the need for models of diffractive scattering which extend to low W and E$_{\pi}$ to be developed and incorporated in simulations.  
Furthermore, these results are useful for understanding the A-dependence of coherent scattering which is important to all oscillation experiments. 

\begin{acknowledgments}

This work was supported by the Fermi National Accelerator Laboratory
under US Department of Energy contract No. DE-AC02-07CH11359 which
included the \minerva\ construction project.  Construction support was
also granted by the United States National Science Foundation under
Award PHY-0619727 and by the University of Rochester. Support for
participating scientists was provided by NSF and DOE (USA), by CAPES
and CNPq (Brazil), by CoNaCyT (Mexico), by CONICYT (Chile), by
CONCYTEC, DGI-PUCP and IDI/IGI-UNI (Peru), and by Latin American
Center for Physics (CLAF).  We thank the
MINOS Collaboration for use of its near detector data. We acknowledge
the dedicated work of the Fermilab staff responsible for the operation and maintenance of the NuMI beamline, MINERvA and MINOS detectors and the physical and software environments that support scientific
computing at Fermilab.

\end{acknowledgments}

\end{document}